\documentclass[prl,twocolumn,a4paper,groupedaddress,showpacs,floatfix,aps,10pt,longbibliography,nobibnotes]{revtex4-2}
\usepackage{graphicx,amsmath,amssymb,amsfonts,dsfont,subfigure,color}
\usepackage{relsize}

\usepackage{color}

\newcommand{\mc}[1]{\ensuremath{\mathcal{#1}}}

\begin{document}

\title{Ion Imaging via Long-Range Interaction with Rydberg Atoms}

\author{Christian Gross${}^{1}$}
\thanks{These two authors contributed equally.}
\author{Thibault  Vogt${}^{1,2}$}
\thanks{These two authors contributed equally.}
\author{Wenhui Li${}^{1,3}$}
\email{wenhui.li@nus.edu.sg}

\affiliation{Centre for Quantum Technologies, National University of Singapore, 3 Science Drive 2, Singapore 117543${}^1$}
\affiliation{MajuLab, CNRS-UCA-SU-NUS-NTU International Joint Research Unit, Singapore 117543${}^2$}
\affiliation{Department of Physics, National University of Singapore, Singapore 117542${}^3$}





\begin{abstract}
We demonstrate imaging of ions in an atomic gas with ion-Rydberg atom interaction induced absorption. This is made possible by utilizing a multi-photon electromagnetically induced transparency (EIT) scheme and the extremely large electric polarizability of a Rydberg state with high orbital angular momentum. We process the acquired images to obtain the distribution of ion clouds and to spectroscopically investigate the effect of the ions on the EIT resonance. Furthermore, we show that our method can be employed to image the dynamics of ions in a time resolved way. As an example, we map out the avalanche ionization of a gas of Rydberg atoms. The minimal disruption and the flexibility offered by this imaging technique make it ideally suited for the investigation of cold hybrid ion-atom systems.
\end{abstract}
\maketitle


The exaggerated polarizibility and dipole moment of a highly excited (Rydberg) atom~\cite{gallagher:ryd} result in its extreme sensitivity to electric fields ranging from static to terahertz (THz) regimes, as well as strong interactions with charged particles and other Rydberg atoms. These properties have enabled applications in the research fields of electric field sensing~\cite{sedlacek2012microwave,holloway2017rfmetrology,Wade2016,PhysRevLett.112.026101}, quantum optics at the single photon level~\cite{dudin2012strongly,peyronel2012quantum,busche2017contactless,tiarks2019photongate}, quantum information processing~\cite{saffmanl2016RydbergQuantumComputing}, and quantum simulation of many-body systems~\cite{bernien2017manybody51,leseleuc2019RydbergTopologica}. Furthermore, the strong interactions between ions and Rydberg atoms are central to the recently observed Rydberg excitation blockade by a single ion~\cite{PhysRevLett.121.193401} and to predictions
on the control of cold atom-ion collisions via Rydberg dressing~\cite{secker2017TrappedIonsRydbergDressed, Ewald2019ObservationOfIonsRydbergInteraction}. These studies are actively pursued in relation to cold hybrid ion-atom systems, which are emerging as a new platform for investigating fundamental quantum physics, including cold collisions, strong coupling polarons, formation of molecular ions, and charge mobilities~\cite{harter2014atomionexp, tomza2019hybridionatom}. The ability to image the ion kinetics \textit{in situ} and non-destructively will be highly relevant for probing such processes.

In this context, electromagnetically induced transparency involving Rydberg states (Rydberg EIT) is promising for imaging the dynamics of ion impurities immersed in cold atom clouds. The underlying idea is to utilize interaction enhanced photon scattering due to the Rydberg excitation blockade~\cite{lukin:01,vogt2007electric,PhysRevLett.121.193401}. A similar concept has been implemented to image impurity Rydberg atoms and monitor dipole-mediated energy transport~\cite{gunter2012interaction,gunter2013observing}. As the imaging technique relies on the absorption of light by a large number of atoms surrounding each ion, it has the potential of being fast and could be applied to ions of any internal structure.


In this letter, we demonstrate imaging of $^{87}$Rb$^+$ ions embedded in a cold cloud of $^{87}$Rb atoms with ion-Rydberg atom interaction induced absorption. We expose the ion-atom mixture to an imaging probe pulse of $ 1 \, \mathrm{\mu s} $ under the conditions of a multi-photon EIT scheme involving the 27$G$ Rydberg state. From the acquired images, we quantitatively analyze the spectral shift and the suppression of the EIT resonance in the presence of ions. Finally, we report a spatially and temporally resolved observation of the avalanche ionization dynamics in a Rydberg gas. This result shows that this imaging technique is well suited to investigate the dynamics of cold hybrid ion-atom systems.

%
\begin{figure}[t!]
\begin{center}
\includegraphics[width=8.5cm]{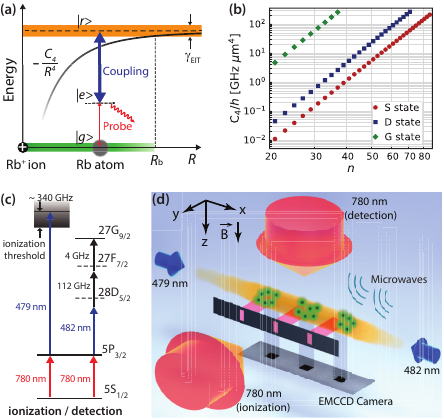}
\end{center}
\caption{\label{fig1}
(a) Concept of the ion-Rydberg atom interaction induced absorption.
(b) Plots of $C_4$ coefficients of the interaction $V_{ir}(R)$ between a $^{87}$Rb$^+$ ion and a $^{87}$Rb Rydberg atom in the $|nG_{9/2}, m_J = 9/2\rangle$ state (green diamonds), the $|nD_{5/2}, m_J = 5/2\rangle$ state (blue squares), and the $|nS_{1/2}, m_J = 1/2\rangle$ state (red dots)~\cite{vsibalic2017arc}.
(c) Photo-ionization and multi-photon Rydberg EIT detection schemes.
(d) Sketch of the experimental setup. A bias $B$ field along $z$ defines the quantization axis. The 479 nm ionization beam is aligned along $x$ with the axial axis of the atomic ensemble. The 780 nm ionization beam passes through a pattern of one or more slits, which is imaged onto the atomic cloud using a $1.8:1$ relay telescope (not shown). The detection fields include a 780 nm probe laser, a 482 nm laser beam counter-propagating with the 479 nm beam, and two microwave fields with wavevectors in the $x-y$ plane. The probe light is scattered inside the Rydberg blockade spheres (green balls) around the ions (black dots), which cast shadows on an electron-multiplying charge-coupled device (EMCCD camera).}
\end{figure}

The principle of the imaging method is illustrated in Fig.~\ref{fig1}(a). A probe beam resonantly drives the atomic transition from the ground state $|g\rangle$ to an excited state $|e\rangle$ while a coupling field is on resonance with the transition from $|e\rangle$ to a Rydberg state $|r\rangle$. In the vicinity of an ion, the energy level $|r\rangle$ is shifted due to the long-range ion-Rydberg atom interaction potential $V_{ir}(R) = - C_4/R^4$, where $R$ is the internuclear distance and $C_4$ is the interaction coefficient. Inside the so-called blockade sphere of radius $R_b= \left( 2 C_4/ \gamma_{EIT} \right)^{1/4}$, this shift exceeds the EIT linewidth $\gamma_{EIT}$, which enhances scattering from state $|e\rangle$. Therefore the atoms within the blockade sphere surrounding an ion absorb the probe light, whereas the rest of the atomic ensemble remains transparent, and this contrast is used to form an image of the ion.

Note that, even in the absence of ions, the transmission of the probe light may be reduced due to the Rydberg-Rydberg interaction $V_{rr}(R) = - C_6/R^6$ ~\cite{pritchard:10,han2016spectral}. Hence the critical aspect to achieve a sufficient imaging contrast is to properly select the Rydberg state $|r\rangle$, such that $V_{ir}(R)$ is strong enough to induce a large absorption around the ion while $V_{rr}(R)$ remains weak. We accomplish this by virtue of the strong dependence of the Rydberg state polarizability $\alpha = 2 (4 \pi \epsilon_0 / e)^2 C_4 $ on the orbital angular momentum $\ell$ of the Rydberg state~\cite{vsibalic2017arc}, where $\epsilon_0$ is the electric constant and $e$ is the elementary charge. As shown in Fig.~\ref{fig1}(b), the $C_4$ coefficient for a $G$ state is two orders of magnitude larger than those for the $S$ and $D$ states of a similar principal quantum number $n$, due to its proximity in energy to the neighboring states of opposite parity. This huge $C_4$ coefficient leads to a blockade radius $R_b > 5 \, \mu$m for a wide range of $\gamma_{EIT}$ in our experiment and for $n$ as low as $n = 20$. A Rydberg state of low $\ell$ and high $n$ may have a $C_4$ coefficient similar to that for a $G$ state of low $n$, but a $C_6$ coefficient three orders of magnitude larger. This $C_6$ would result in a nearly complete absorption of the EIT probe light under the experimental conditions necessary for imaging~\cite{han2016spectral}. In balancing the two interaction effects as well as taking into account technical factors in our setup, we have chosen the state $|r_0\rangle = |27G_{9/2}, m_J = 9/2\rangle$ as the upper EIT level, which we access via a multi-photon EIT scheme~\cite{vogt2018microwave}.


We begin our experiment with the preparation of an ensemble of $^{87}\textrm{Rb}$ atoms in the ground state  $|g\rangle = |5S_{1/2}, F = 2, m_F = 2\rangle $, which is confined in a single-beam optical dipole trap (ODT) at a temperature of $  25 \, \mu \mathrm{K} $. After being released from the ODT and a time-of-flight (TOF) of $ 10 \, \mathrm{\mu s} $, the highly elongated atomic cloud has a peak density of $ 4.8\times10^{11} \, \mathrm{cm}^{-3} $ and radially follows a Gaussian density distribution with a $1/e^2$ radius $w = 12 \, \mu \mathrm{m}$. To produce ions, we use the two-photon ionization scheme shown in Fig.~\ref{fig1}(c), where a 780 nm laser resonantly drives the transition to the intermediate state $|e\rangle = |5P_{3/2}, F = 3, m_F = 3\rangle $ and a 479 nm laser provides the coupling to the continuum at about $ 2\pi\times 340 \, \mathrm{GHz} $ above the ionization threshold. The photo-ionization lasers are applied for a duration of $t_i = 0.14 \, \mu\mathrm{s} $ to generate ions within  well-defined regions of the atomic ensemble, as illustrated in Fig.~\ref{fig1}(d).

Following the two-photon ionization pulse and a delay of $0.1 \, \mu\textrm{s}$, we perform imaging with a $ 1 \, \mu \mathrm{s} $ EIT detection pulse. As shown in Fig.~\ref{fig1}(c), the probe field of wavelength $\lambda_P = $ 780 nm couples to the $|g\rangle \rightarrow |e\rangle$ atomic transition. An effective three-photon coupling field is used to enable the $|e\rangle \rightarrow |r_0\rangle $ transition, via the off-resonant intermediate states $ |28D_{5/2}, m_J = 5/2\rangle $ and $ |27F_{7/2}, m_J = 7/2\rangle $. It is formed by the optical field D (482 nm) and the two microwave fields M1 (112 GHz) and M2 (4 GHz). The spatial configuration of the imaging fields is shown in Fig.~\ref{fig1}(d). The circularly polarized probe beam of peak Rabi frequency $ \Omega_P $ = $2\pi \times 0.7 \, \mathrm{MHz} $  propagates through the atomic cloud along the quantization axis $z$, and is projected onto the EMCCD camera using an imaging system of $3 \, \mu\textrm{m}$ resolution~\cite{han2015lensing}. We choose a magnetic field $B$ of 6.1 G to induce sufficiently large Zeeman splittings of the energy levels, and appropriate frequencies and helicities of the fields D, M1 and M2, such that only the $\sigma^+$ transitions are of relevance in our experiment. These three fields form an effective coupling field, with experimentally calibrated Rabi frequency $\Omega_{C,eff}=2\pi \times 6.7$~MHz. Further experimental details on the imaging fields are given in Ref.~\cite{supp}.

\begin{figure}[t!]
\begin{center}
\includegraphics[width=8.5cm]{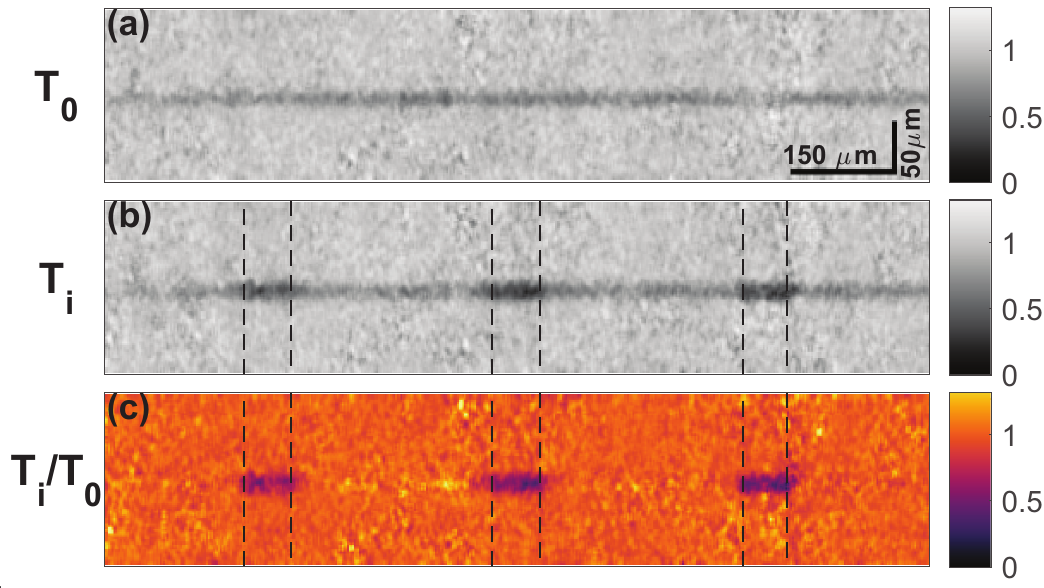}
\end{center}
\caption{\label{fig2}
Imaging of ion clouds. (a) EIT image $T_0$ acquired without ions. (b) EIT image $T_i$ acquired in presence of three ion clouds. (c) Image of the relative transmission $T_i/T_0$. Each of the images (a) and (b) is the average of about 220 preprocessed images. The dashed lines delimit the areas where the atoms are exposed to the photo-ionization 780 nm laser beam. About 11 ions are created in each of the three exposed areas.
}
\end{figure}

The obtained images are preprocessed to remove undesirable interference fringes~\cite{li2007reduction}, and then averaged over many experimental cycles. In the absence of the ionization pulse, we observe an approximately uniform absorption along the atomic cloud under resonant EIT conditions, as shown in Fig.~\ref{fig2}(a). This residual absorption is due to the Rydberg blockade induced dissipation resulting from the interaction $V_{rr}(R)$~\cite{pritchard:10,han2016spectral}. It could be reduced by lowering $ \Omega_P $, which however would be detrimental to the imaging signal-to-noise ratio. When an ionization pulse is applied in three different regions of the atomic cloud, as indicated by dashed lines in Fig.~\ref{fig2}(b), an enhanced absorption is observed at the positions of the ion clouds. We extract the image of ions by computing the relative transmission at every pixel position~$(x,y)$,
\begin{equation}
\mc{T}(x,y) = \frac {T_i(x,y)}{T_{0}(x,y)},
\label{relative transmission}
\end{equation}
where $T_i$ and $T_0$ correspond to the transmission images acquired with and without applied ionization pulse, respectively. This post-processing highlights the local character of the ion distribution, as shown in Fig.~\ref{fig2}(c).

%

\begin{figure}[t!]
\begin{center}
\includegraphics[width=8.5cm]{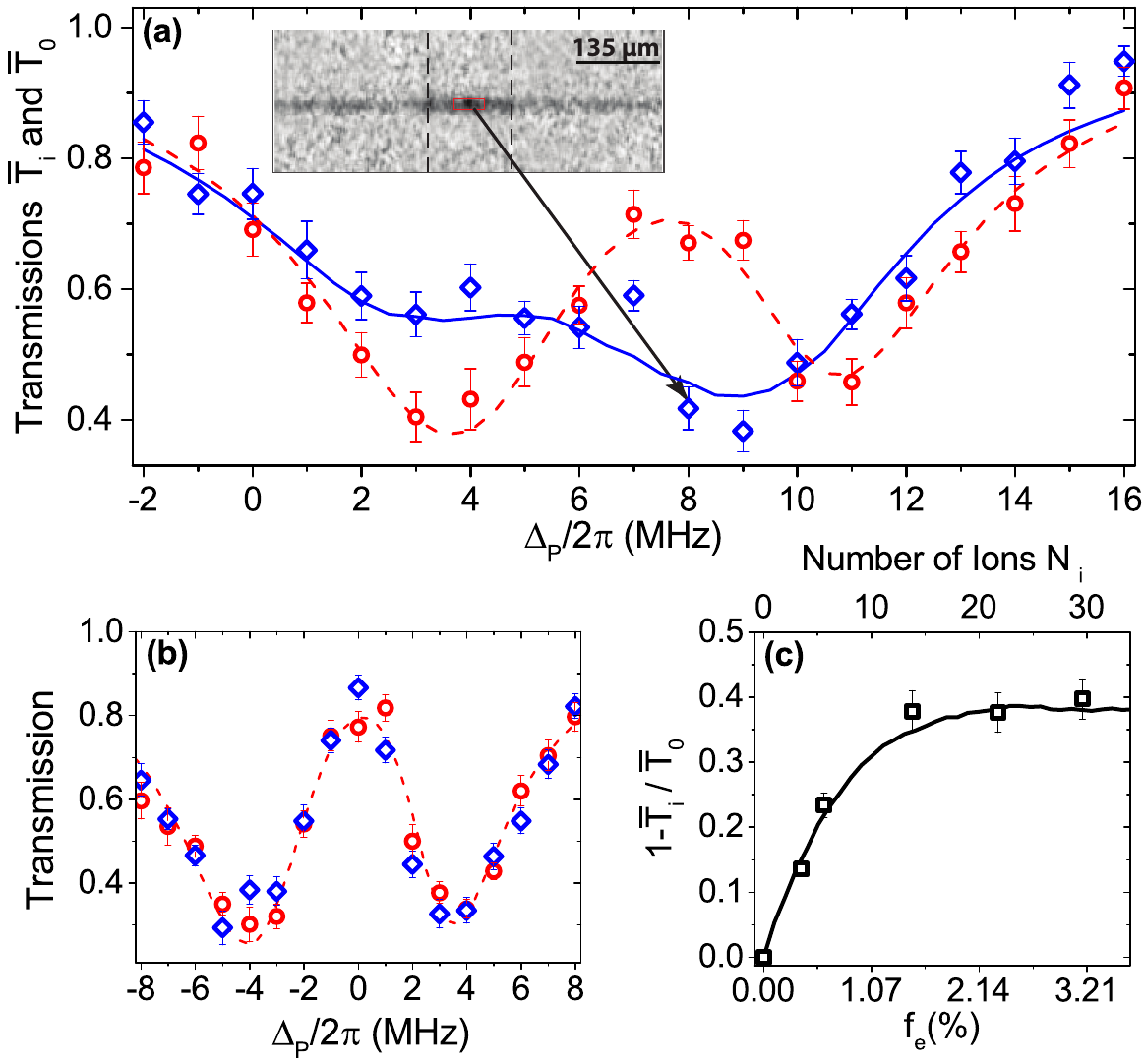}
\end{center}
\caption{\label{fig3}
(a) and (b) EIT spectra showing mean probe transmissions $\overline{T}_i$ and $\overline{T}_0$ vs. probe field detuning $\Delta_P$ using the Rydberg state $|27G_{9/2}, m_J = 9/2\rangle$ in (a) and $|28D_{5/2}, m_J = 5/2\rangle $ in (b). The data markers correspond to experimental spectra recorded without ions (red circles) and with ions (blue diamonds). The EIT image $T_i$ shown in the inset is the average of 22 preprocessed images.  (c) Relative absorption $\mc{A}=1-\overline{T}_i/\overline{T}_0$ recorded for $\Delta_{P}=2 \pi \; \times 8 \ \textrm{MHz}$ as a function of $f_e$ and $N_i$. The dashed lines are 3-level EIT fits, and the solid lines are the result of a theoretical model as detailed in the text.  The error bars are the standard error over more than 20 imaging realizations.
}
\end{figure}

The images of Figs.~\ref{fig2}(a) and \ref{fig2}(b) are acquired under optimized conditions based on the spectroscopic studies shown in Fig.~\ref{fig3}. For the measurements in Fig.~\ref{fig3}, we apply a photo-ionization pulse in a single region of the atomic cloud and vary the probe field frequency detuning $ \Delta_P $ at the beginning of each experimental cycle, while keeping all other parameters unchanged. The two four-photon EIT spectra in Fig.~\ref{fig3}(a) correspond to the mean transmissions $\overline{T}_i$ and $\overline{T}_0$ recorded with and without ionization pulse, respectively. Here, $\overline{T}_i$ and $\overline{T}_0$ are calculated over the small region-of-interest (ROI) indicated by a rectangle in the inset image of Fig.~\ref{fig3}(a).

In the absence of the ionization pulse, the spectrum is similar to those of standard Rydberg EIT configurations~\cite{fleischhauer:05}, as confirmed by a fit with the steady-state solution of a 3-level EIT system (Fig.~\ref{fig3}(a)). We observe a characteristic transmission peak centered in the broad atomic absorption spectrum, which, however, is shifted from the atomic transition frequency by $\Delta_{P0} \approx 2 \pi \times 8\; \textrm{MHz}$. This shift is due to AC Stark shifts coming from the off-resonant fields that form the effective coupling field~\cite{vogt2018microwave}. In the presence of the ionization pulse, generating approximately 14 ions, the drop in transmission is sharp around $\Delta_{P0}$. This drop is due to interaction enhanced scattering within the blockade sphere of radius $R_b$ around each impurity ion. Using the ion-atom interaction coefficient for $|r_0\rangle$ given in Fig.~\ref{fig1}(a), $C_4/h=26.5\; \textrm{GHz}\, \mu \textrm{m}^4 $, we obtain $R_b = 8.5\; \mu \textrm{m}$, which is comparable to the radius of the atomic cloud. This large blockade radius is consistent with the disappearance of the EIT effect at $\Delta_{P0}$. Moreover, the EIT resonance is red-shifted by about $2 \pi \times 2.5\; \mathrm{MHz}$ from $\Delta_{P0}$, while being broadened and strongly attenuated. This is due to the attractive and inhomogeneous ion-Rydberg atom interaction in the present system.

The large $C_4$ coefficient of state $|r_0\rangle$ is essential for this imaging technique. To verify this, we change the upper EIT state from $|r_0\rangle$ to the nearby $|28D_{5/2}, m_J = 5/2\rangle$ state having a $C_4$ coefficient two orders of magnitude smaller. We achieve this with a standard two-photon EIT scheme, where the coupling field is provided by a single 482 nm laser. The corresponding EIT spectra acquired with and without ionization pulse are shown in Fig.~\ref{fig3}(b), and no clear difference is noticeable. This is expected, as the blockade radius in this case is reduced threefold, which means that $\sim$ 30 times fewer probe photons are scattered off around an ion.

%
%

In Fig.~\ref{fig3}(c), we report on the relative absorption $\mc{A}=1-\overline{T}_i/\overline{T}_0$ as a function of the fraction of atoms excited to the intermediate state $ |e\rangle $, $f_e$, which is varied using the intensity of the 780 nm ionization laser. The total number of ions $N_{i}$ in the ionization volume depends linearly on $f_e$, and is given on the top horizontal axis of Fig.~\ref{fig3}(c). The relative absorption first increases steadily with $N_{i}$ and then saturates rapidly above $N_{i} \sim 10$. For small $N_i$ where the blockade spheres surrounding the ions are unlikely to intersect, $\mc{A} \propto N_{i}$, as explained in~\cite{supp}. When $N_{i}$ increases, the blockade spheres start to overlap and the number of scattered photons per ion is reduced, resulting in the saturation of $\mc{A}$. In this regime, the relative absorption reaches $\mc{A}_{sat} \approx$ 0.4, which defines the maximum contrast achieved with our current experimental conditions.
We infer that a contrast of $\sim 0.3$ or more can be expected from a single blockade sphere of radius $R_b$ induced by an individual ion.

To obtain a more quantitative understanding of our experimental results, we compare the data of Fig.~\ref{fig3} with the model presented in Refs.~\cite{vogt2018microwave,supp}. The propagation of the probe field $\mc{E}_P$ through the atomic cloud is calculated in steady-state using the differential equation $\partial_{z}\mc{E}_P(\vec r)=  i  \pi /\lambda_P \; \chi(\vec r) \mc{E}_P(\vec r)$, where $\chi(\vec r)$ is the linear susceptibility at position $\vec r $. For our effective three-level system, $\chi(\vec r)$ is approximated as

\small
\begin{align}
\chi \left ( \vec r \right)&= \frac{i\, n_{at} \left( \vec r \right) \Gamma \sigma \lambda_P/ 4\pi}{\frac{\Gamma}{2}-i (\Delta _{P} - \delta_{1})+\frac{\Omega _{C,eff}^2 /4}{ \gamma -i \left[\Delta _P + \Delta_{C} - \delta_2 - \delta_{Stark}\left( \vec r \right)\right]}}
\label{susceptibility},
\end{align}
\normalsize
where $\Gamma/ 2 \pi= 6.067\; \mathrm{MHz}$ is the spontaneous decay rate from state $|e \rangle$, $n_{at}\left( \vec r \right)$ is the atomic density, $\gamma$ is a dephasing rate of the atomic coherence between the $|g\rangle$  and $|r_0\rangle$ states, and $\sigma$ is a scattering cross-section~\cite{supp}. Moreover, $\Delta_{C}$ is the three-photon detuning from the $|e \rangle \rightarrow |r_0\rangle$ atomic transition, and $\delta_{1}$ and $\delta_{2}$ are the AC Stark shifts affecting the $|e \rangle$ and $|r_0\rangle$ energy levels, respectively. Finally, $\delta_{Stark}\left( \vec r \right)=\alpha(\theta) \, \mathbf{E}\left( \vec r \right)^2 / 2$ is the Stark shift of $|r_0\rangle$, where $\alpha(\theta)$ is the polarizability depending on $\theta$, the angle between the quantization axis and the direction of the electrostatic field $\mathbf{E}$ at the position $\vec r$~\cite{supp}.

In the simulation, each ion sample is generated given the ionization rate per atom $R_{I}\left( \vec r \right)=\varrho_{ee}\left( \vec r \right) \sigma_I \Phi_{479}$, where $\sigma_I$ is the ionization cross-section at the wavelength of the 479 nm laser, $\Phi_{479}$ is its photon flux density, and $\varrho_{ee}$ is the fractional excitation in the intermediate state $ |e\rangle $. The consideration of the ion cloud expansion during the detection pulse, the effect of which may not be negligible for a large ion number, is discussed in Ref.~\cite{supp}. The simulated curves displayed in Fig.~\ref{fig3} are averages over many ion samples and calculated for $\sigma_I=20.4\; \textrm{Mb}$, which is consistent with the cross section reported in Ref.~\cite{Nadeem2011photoionization}. With all the other parameters being experimentally calibrated, the good agreement between the data and the simulation confirms that the ion-Rydberg atom interaction is the sole mechanism responsible for the observed imaging contrast.


\begin{figure}[t!]
\begin{center}
\includegraphics[clip,width=8.5cm]{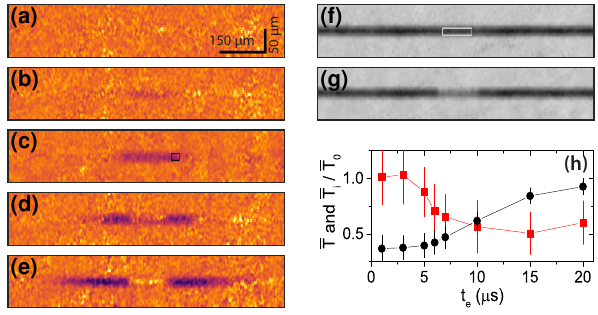}
\end{center}
\caption{\label{fig4}
Avalanche ionization. Interaction induced images of the space charges after an excitation duration $t_e = 0.14\; \mu \mathrm{s}$ (a), $3\; \mu \mathrm{s}$ (b), $5\; \mu \mathrm{s}$ (c), $7\; \mu \mathrm{s}$ (d), and $10\; \mu \mathrm{s}$ (e). Corresponding absorption images of ground-state atoms acquired after $t_e = 5\; \mu \mathrm{s}$ (f) and $10\; \mu \mathrm{s}$ (g). (h) Relative transmission ($\overline{T}_i/\overline{T}_0$) extracted from the ion images at the position just outside the excitation region (red squares) and transmission ($\overline{T}$) of the absorption images at the center of the excitation region (black dots) vs $t_e$. The boxes on (c) and (f) indicate the areas for extracting $\overline{T}_i/\overline{T}_0$ and $\overline{T}$, respectively.
}
\end{figure}
%
%

Our ion imaging scheme opens up the possibility of spatially investigating ionization processes and the dynamics of ion-atom hybrid systems. We illustrate this in Fig.~\ref{fig4} with a series of images showing the avalanche ionization of an atomic gas excited to the $|30D_{5/2}, m_J = 5/2\rangle $ state. For these measurements, the 479 nm ionization laser in Fig.~\ref{fig1} is replaced by a 482 nm laser for the coupling to the Rydberg state, and the excitation is performed with a single slit for a variable duration $t_e$. After a delay of $0.1 \, \mu\textrm{s}$, we either monitor the ion distributions with our imaging scheme in Figs.~\ref{fig1}(c) and \ref{fig1}(d) or take absorption images of the ground-state atoms. The images in Fig.~\ref{fig4} are consistent with the well established avalanche ionization process~\cite{PhysRevLett.110.045004,PhysRevA.86.020702}, and this is confirmed by the presence of a time threshold in the transmission plots of Fig.~\ref{fig4}(h). While no ions are observed at $t_e = 0.14\; \mu \mathrm{s}$ (Fig.~\ref{fig4}(a)), an initial ionization occurs on the scale of a few microseconds, mostly due to Penning ionization of pairs of cold Rydberg atoms excited to the $|30D_{5/2}, m_J = 5/2\rangle $ state~\cite{li:05}. At this stage, the ion distribution remains localized at the excitation region (Figs.~\ref{fig4}(b) and ~\ref{fig4}(c)), and these few ions barely affect the distribution of the ground-state atoms (Fig.~\ref{fig4}(f)). Once the space charge due to the ions is sufficiently large to trap electrons, the avalanche ionization is set off by electrons colliding with Rydberg atoms and a plasma forms. This process rapidly depletes the ground-state atoms upon continuous coupling the atomic cloud to the Rydberg state (Fig.~\ref{fig4}(g)). With this ion imaging technique, it can be clearly seen in Figs.~\ref{fig4}(d) and ~\ref{fig4}(e) that the plasma expands into the atomic cloud outside the excitation region. We note that the probability of ionizing the atoms in state $|r_0\rangle$ during the EIT imaging pulse of 1~$\mu$s remains small in the presence of the plasma, as $\Omega_P$ is low and the fraction of atoms in $|r_0\rangle$ is further suppressed due to the large Stark shifts. Thus our imaging technique can image the space charge and its expansion with minimum disturbance added to the plasma.


In conclusion,  we have demonstrated \textit{in situ} imaging of ions in an atomic ensemble by ion-Rydberg atom interaction induced absorption, with which we have visualized avalanche ionization processes. One advantage of this imaging technique is that it does not depend on the internal structure of the ionic impurities and can be applied to image a wide range of ion-atom mixtures, including ion species with no cycling optical transitions available. Our technique is currently limited by the low intensity of the probe light, which is used to avoid excess absorption due to Rydberg-Rydberg interactions but prevents us from single-shot imaging. A homodyne detection of the probe light with a reference beam of a ten times stronger field strength would sufficiently improve the signal-to-noise ratio~\cite{kadlecek2001nondestructive} for spatially resolving individual ions in a single shot.

\section*{Acknowledgement}
\begin{acknowledgments}
The authors thank Tom Gallagher and Martin Kiffner for helpful discussions and acknowledge the support by the National Research Foundation, Prime
Ministers Office, Singapore and the Ministry of Education, Singapore under the Research Centres of Excellence programme.
\end{acknowledgments}

%
%

\clearpage
\pagebreak
\begin{titlepage}
\begin{center}
\section*{Supplementary material: Ion Imaging via Long-Range Interaction with Rydberg Atoms}
\end{center}
This supplementary material provides additional experimental details. It also presents our model for simulating the transmission of a probe light through an atomic cloud driven in configuration of Rydberg EIT in the presence of ions. The first section gives tables of experimental parameters, the second section discusses the simulation of the ion cloud, the third section focuses on the propagation of the probe light, and finally the fourth section discusses the scaling of the transmission with the number of ions in the cloud to be imaged. \\
\end{titlepage}

\setcounter{secnumdepth}{2}
\setcounter{equation}{0}
\setcounter{figure}{0}
\setcounter{table}{0}
\setcounter{page}{1}



\section{Relevant experimental parameters}
\subsection{Ionization beams}

A two-photon ionization scheme, shown in Fig. 1(c) of the main text, is used for producing ions in the atomic ensemble. Here is a list of all relevant parameters in regard to the two ionization beams, in reference to the experimental configuration shown in Fig. 1(d) of the main text. Also listed are the parameters of the Rydberg excitation beams used for acquiring the data shown in Fig. 4 of the main text.

\small
\begin{table}[!h]
\caption{\label{tab:DipolesIntensities} Experimental parameters related to the two ionization (excitation) laser beams }
			\begin{ruledtabular}
			\begin{tabular}{ll}
			  Parameters    & Value / Unit \\
                    \hline
              of the 780 nm ionization (excitation) beam  &    \\
			 			  \hline			
		  	 Propagation direction    & $-\hat{y}$   \\
				Polarization     & $\hat{x}$  \\
				$1/e^2$ radius (before the slits) & $ 3.5 $ mm \\
				Rabi frequency ($\sigma^+$ component) & 0 - 1.5 (1.8)  MHz \\
                    \hline
              of the 479 nm ionization beam  &    \\   	
                    \hline	
                Propagation direction  & $\hat{x}$  \\
				Polarization   & $\hat{y} $   \\
				$1/e^2$ radius & $ 31 $ $\mu$m \\
				Power  & 137 mW  \\
                 \hline
              of the 482 nm excitation beam  &    \\   	
                    \hline	
                Propagation direction  & $\hat{x}$  \\
				Polarization   & $\hat{y} $   \\
				$1/e^2$ radius & $ 31 $ $\mu$m \\
				Rabi frequency ($\sigma^+$ component) & 7.5  MHz
			\end{tabular}
			\end{ruledtabular}
\end{table}
\normalsize

\subsection{Detection fields}

A multi-photon Rydberg EIT scheme, shown in Fig. 1(c) of the main text, is used for imaging ions in the atomic ensemble. Here is a list of all relevant parameters in regard to the probe beam and the fields D, M1, M2 that form the effective coupling field, in reference to the experimental configuration shown in Fig. 1(d) of the main text.

\small
\begin{table}[!h]
\caption{\label{tab:DipolesIntensities} Experimental parameters related to the detection fields }
			\begin{ruledtabular}
			\begin{tabular}{ll}
			  Parameters    & Value / Unit \\
                    \hline
              of the 780 nm probe beam  &    \\
			 			  \hline			
		  	 Propagation direction    & $+\hat{z}$   \\
				Polarization     & $\hat{\sigma}^+$   \\
				$1/e^2$ radius & $ 3.4 $ mm \\
				Rabi frequency $\Omega_P$ &  $2 \pi \times 0.7\; \mathrm{MHz}$  \\
                    \hline
              of the 482 nm  beam (D field) &    \\   	
                    \hline	
                Propagation direction  & $-\hat{x}$   \\
				Polarization   & $\hat{y} $   \\
				$1/e^2$ radius & $ 46.5 \, \mu$m \\
                Frequency detuning $\Delta_D$  & $ 2 \pi \times 32 \, \mathrm{MHz} $ \\
				Rabi frequency ($\sigma^+$ component) $\Omega_D$  & $ 2 \pi \times 34 \, \mathrm{MHz} $ \\
                     \hline
              of the fields M1 and M2  &    \\   	
                    \hline	
                Propagation direction  & in the $\hat{x}-\hat{y}$ plane   \\
				Polarization   & in the $\hat{x}-\hat{y}$ plane   \\
		        Frequency detuning $\Delta_{M1}$  & $2 \pi \times  \, 60 \, \mathrm{MHz} $ \\
				Frequency detuning $\Delta_{M2}$ & $- 2 \pi \times  \, 81.5\, \mathrm{MHz} $ \\
                Rabi frequency ($\sigma^+$ component) $\Omega_{M1}$ & $2 \pi \times  \, 49\, \mathrm{MHz} $ \\
				Rabi frequency ($\sigma^+$ component) $\Omega_{M2}$ & $2 \pi \times  \, 50\, \mathrm{MHz} $
			\end{tabular}
			\end{ruledtabular}
\end{table}
\normalsize

\section{Simulation of the ion cloud}
\subsection{Excitation scheme: Reminder}
Rubidium atoms in an atomic cloud are photo-ionized with laser pulses applied for a duration of $t_i = 140 \; \textrm{ns}$ using a two-step photo-ionization scheme. The first step is realized with a 780 nm laser ($\lambda_{780}=780.228\ \textrm{nm}$) resonant with the $|5S_{1/2}, m_F=2\rangle \leftrightarrow |5P_{3/2}, m_F=3\rangle $ atomic transition and of intensity $I_{780}$. For the second step, we use a 479 nm laser ($\lambda_{479}=478.799\ \textrm{nm}$) tuned above the ionization limit by about 340 GHz and of intensity $I_{479}$. The 340 GHz detuning is used to ensure that the photo-ionized electrons escape rapidly from the atomic cloud. The intensity $I_{479}$ is approximately uniform on the atomic cloud and $I_{780}$ is uniform along $x$ within the limits set by the pattern mask used in the experiment. We will only consider the case of a single slit mask for simplicity. The Coulomb expansion during the photo-ionization time of $t_i$ is neglected and the atomic density is quasi-uniform along $x$, thus the ion density obtained after photo-ionization follows a rectangle function along $x$. It varies radially along $y$ and $z$ due to the atomic density variation and the absorption of the 780 nm light. For convenience with the notations, we will assume that the reference frame has been \it{rotated by $\pi$ \normalfont around the $z$ axis compared to that given in Fig. 1(d) of the main text, such that the 780 nm light propagates \it{along the $y$ axis instead of the $-y$ axis}.\normalfont

\subsection{Photo-ionization rate}
The ionization rate of an atom at position $\vec r=(x,y,z)$ inside the atomic cloud is
\begin{equation}
R_{I}(\vec r)=\rho_{ee}(\vec r) \frac{\sigma_I I_{479} \lambda_{479}}{h \, c},\label{IoizationRate}
\end{equation}
where $\rho_{ee}$ is the fractional excitation in the intermediate state $|5P_{3,2}, m_F=3\rangle$, $h$ is the Planck constant, $c$ is the speed of light in vacuum, and $\sigma_I$ is the ionization cross-section at the wavelength $\lambda_{479}$ used in the experiment. The ionization cross-section has been measured in Ref.~\cite{Nadeem2011photoionization} as $\sigma_I=18.8\pm 3.8 $~Mb (1 barn = $10^{-28}\ \textrm{m}^{2}$).
On resonance, $\rho_{ee}$ is given by
\begin{equation}
\rho_{ee} = \frac{1}{2}\frac{s}{1 + s },
\end{equation}
where the saturation parameter $s$ equals
\begin{equation}
s(x,y,z)=\frac{I_{780}(x,y,z)}{I_{sat}},
\end{equation}
given the saturation intensity $I_{sat}=16.69 \, \mathrm{W/m}^2$.

For simplicity, we will assume that the input 780 nm beam is collimated and uniform in the $z$ direction on the scale of the atomic cloud, with input saturation parameter $s_0(x)\approx s(x,-\infty,0)$.
The saturation parameter is the solution of the differential equation
\begin{equation}
\frac{d s}{d y} = -  \sigma_0 n_{at}\frac{s}{1 + s }, \label{equadifSaturationParameter}
\end{equation}
where $\sigma_0=3 \lambda_{780}^2 /2\pi$ is the resonant cross-section, $n_{at}$ is the atomic density. The exact solution of Eq. \eqref{equadifSaturationParameter} is known and can be expressed with the product logarithm function:
\begin{equation}
s(x,y,z)=\textrm{ProductLog}\left[s_0(x)\, \mathrm{e}^{-(D_P (y,z) - s_0(x))}\right],
\end{equation}
where the optical depth $D_P$ reads
\begin{equation}
D_P(y,z) =\int_{-\infty}^y \textrm{d}y' \sigma_0  n_{at}(y',z).
\end{equation}
\subsection{Rejection sampling algorithm}
Knowing the ionization rate at position $\vec r =(x,y,z)$ in the sample (see Eq. \eqref{IoizationRate}), we generate typical ion cloud samples obtained after the photo-ionization time of $t_i$. The ion density at position $\vec r$ after excitation is
\begin{equation}
n_{i}(\vec r)=  n_{at}(\vec r) R_I \left( \vec r\right) t_i,
\end{equation}
and the total ion number is $N_i=\int \mathrm{d }\vec r n_i(\vec r)$.
To generate a numerical simulation of the ion cloud, the ions must be sampled according to the multivariate distribution $\mathbf{R}=(X,Y,Z)$ with joint probability density
\begin{equation}
f_{\mathbf{R}}(x,y,z)=\frac{ n_{i}(\vec r)}{N_i}.
\end{equation}
A simple sampling algorithm for simulating this type of distribution is rejection sampling. It consists in calculating a known distribution $g$ which is sufficiently close to the desired distribution and which must fulfill the condition $g(x,y,z) A \geq  f_{\mathbf{R}}(x,y,z)$ for every point $(x,y,z)$ with $A$ being a constant.

To apply this algorithm, we consider the distribution $\mc{N}$ obtained by neglecting any 780 nm light absorption in the atomic cloud. If $I_{780}$ were constant in the sample, the corresponding ionization rate $R_{I,0}$ would be uniform and the probability density $g(x,y,z)$ would be Gaussian shaped along $y$ and $z$ as it would be simply proportional to $n_{at}$
\begin{equation}g(x,y,z)=\frac{ n_{i,0}(\vec r)}{N_{i,0}}=\frac{ n_{at}(x,y,z)}{N_0},
\end{equation}
where $n_{i,0}(\vec r)=  n_{at}(\vec r) R_{I,0} t_i$ is the ion density in this ideal case, $N_{i,0}=\int \mathrm{d }\vec r \, n_{i,0}(\vec r)$, and $N_0$ is the total atom number in the region illuminated by the 780 nm light.
It is obvious that the inequality $n_{i,0}(x,y,z)\geq n_{i}(x,y,z)$ holds for every point $(x,y,z)$ and therefore
\begin{equation}
g(x,y,z) \times  A  \geq f_{\mathbf{R}}(x,y,z),
\end{equation}
where $A=N_{i,0}/N_i$.

\noindent Then, the rejection sampling algorithm is implemented as follows
\begin{itemize}
	\item Obtain an ion location $(x_i,y_i,z_i)$ according to the distribution $\mc{N}$.
	\item Sample a random number $u$ from a uniform distribution between 0 and 1.
	\item Accept the sample $(x_i,y_i,z_i)$ at the condition that $u< \frac{f_{\mathbf{R}}(x_i,y_i,z_i)}{A g(x_i,y_i,z_i)}$.
	\item Run the algorithm until the wanted number of ions $N_i$ has been reached.
\end{itemize}

\noindent The total number of ions is assumed to be fixed to $N_{i}$ in each sample of the simulation. Ideally, one would also sample the total ion number $N_i$ according to a Poisson distribution. It is assumed that this fluctuation of $N_i$ has only a minor effect on simulated average images.

\subsection{Ion cloud expansion}
\label{expansion}

In the previous subsection, we have presented how to numerically generate a random sample of ions, while neglecting the Coulomb dynamics during the ionization time of $t_i$. The expansion of the ion cloud is not negligible for longer durations and large ion numbers and thus must be taken into account during the detection stage. In this subsection, we describe how we simulate the expansion of a cloud of ions.

In the cloud of Rb$^+$ ions, the force acting on one ion $i$ due to the presence of the other ions $j$ is
\begin{equation}
\overrightarrow{F_i}=\sum_{j \neq i} \frac{q^2 \overrightarrow{r_{ji}}}{4 \pi \epsilon_0 r_{ji}^3},
\end{equation}
where $\overrightarrow{r_{ji}}=\overrightarrow{r_i} -\overrightarrow{r_j}$, $r_{ji}$ is the distance between the ions $i$ and $j$, $q$ is the unit charge and $\epsilon_0$ is the vacuum permittivity.
\noindent According to Newton's laws of motion, the expansion of the cloud is the solution of the following system of equations:
\begin{align}
\frac{d\overrightarrow{v_i}}{dt}&= \frac{\overrightarrow{F_i}} {m_i}, \\
\frac{d\overrightarrow{r_i}}{dt}&= \overrightarrow{v_i},
\end{align}
where $m_i \equiv M_{Rb^+}$ is the mass of a Rubidium ion.
In our numerical simulation, this system of equations is solved using Euler's method. The time is discretized and the distribution of positions $\overrightarrow{r_i}$ and velocities $\overrightarrow{v_i}$ are computed at each time step $\delta t$:
\begin{align}
\overrightarrow{v_i}(t+\delta t)&= \overrightarrow{v_i}(t) + \frac{\overrightarrow{F_i}(t)}{m_i} \, \delta t  \\
\overrightarrow{r_i}(t+\delta t)&= \overrightarrow{r_i}(t)+\overrightarrow{v_i}(t) \, \delta t
\end{align}
A sufficiently small time step $\delta t$ is chosen such that the trajectories of close pairs of ions are well simulated.

\section{Calculating the probe field transmission in presence of ions}

\subsection{Five-level G-EIT and effective three-level G-EIT}

The detection method used for imaging makes use of the EIT scheme described in the main text. The probe field of Rabi frequency $\Omega_P$ addresses the transition $|g\rangle \leftrightarrow |e\rangle$ while an effective coupling field drives the transition $|e\rangle \leftrightarrow |r_0\rangle$ via the off-resonant intermediate states $ |28D_{5/2}, m_J = 5/2\rangle $ and $ |27F_{7/2}, m_J = 7/2\rangle $. We remind that $|g\rangle \equiv |5S_{1/2}, m_F=2\rangle$, $|e\rangle \equiv |5P_{3/2}, m_F=3\rangle $, and $|r_0\rangle \equiv |27G_{9/2}, m_J=9/2\rangle$. The effective field is realized with the three fields D, M1, and M2 of Rabi frequencies $\Omega_D$, $\Omega_{M1}$, and $\Omega_{M2}$, detuned from their corresponding atomic transitions by $\Delta_D$, $\Delta_{M1}$, and $\Delta_{M2}$, respectively.

The propagation of the probe field through the atomic cloud is simulated in steady state within the framework of Maxwell-Bloch' equations in rotating wave, slow envelope, and local density approximations, while neglecting lensing and diffraction effects as well~\cite{han2018coherent}.
With these approximations, the Rabi frequency of the probe field satisfies the following differential equation
\begin{align}
\partial_{z}\Omega_{p}(\vec r)=  i k_P \frac{\Omega_p}{2}\chi(\vec r),
\label{maxwell}
\end{align}
where $k_P = 2 \pi / \lambda_{P}$, and $\chi(\vec r)$ is the susceptibility at position $\vec r $, expressed at first order in $\Omega_P$.
The susceptibility is approximated by that of an effective three-level system
\begin{align}	
\chi \left ( \vec r \right)& \approx \frac{i \, n_{at} \left( \vec r \right) \Gamma \sigma / k_P}{ \Gamma-i 2 \Delta'_{P}+\frac{\Omega _{C,eff}^2}{2 \left[\gamma-i(\Delta _P+\Delta'_{C}-\delta_{Stark})\right]}} \label{susceptibility},
\end{align}%
where $\gamma$ is a dephasing rate of the atomic coherence between states $|g\rangle$ and $|r_0\rangle$, and $\sigma$ is a scattering cross-section. The effective detunings $\Delta'_{P}=\Delta_P -\delta_1$ and $\Delta'_{C}=\Delta_D+\Delta_{M1}+\Delta_{M2}-\delta_2$ include the AC Stark shifts $\delta_1$ and $\delta_2$ of the levels $|5P_{3/2}, m_F=3 \rangle$ and $|27G_{9/2},mJ=9/2\rangle$, respectively. The Stark shift $\delta_{Stark}$ due to the ions can be calculated from the  polarizability of state $|r_0\rangle$ as detailed in the next subsection.
The parameters $\sigma$, $\delta_1$, $\delta_2$, $\Omega_{C,eff}$, and $\gamma$ are obtained from a fit to the spectrum acquired in the absence of ions.

\subsection{Energy level shifts due to Stark effect}

In the presence of an ion, the energy level $|r_0\rangle$ of surrounding atoms is locally shifted by Stark effect.
The electrostatic field at the position $\overrightarrow{r_k}$ of an atom and due to the presence of ions located at positions $\overrightarrow{r_j}$  is
\begin{equation}
\overrightarrow{\mc{E}}(\vec r_k)=\sum_{j} \frac{q \ \ \overrightarrow{r_{jk}}}{4 \pi \epsilon_0 \, r_{jk}^3}.
\end{equation}
The stark shift is approximated as
\begin{equation}
\Delta E_k=\frac{1}{2} \alpha \left( \theta_k \right) \mc{E}^2(\vec r_k),
\end{equation}
where $\alpha$ is the polarizability of energy level $|r_0\rangle$ for a given angle $\theta_k$ between the quantization axis and the axis of the electrostatic field.
The polarizability is extracted from the Stark diagram calculated with the \textit{pairinteraction} software~\cite{weber2017calculation}. A fit to the extracted polarizabilities shown in Fig.~\ref{polarizability} yields the following good heuristic approximation
\begin{equation}
\alpha \left( \theta \right)=\alpha_0+\alpha_2 \sin^4\theta
\label{polarEq}
\end{equation}
where $\alpha_0=-270\; \textrm{MHz}\, \textrm{V}^{-2}\, \textrm{m}^{2}$ and $\alpha_2=-170\; \textrm{MHz}\, \textrm{V}^{-2}\, \textrm{m}^{2}$ for the $|27G_{9/2}, m_J=9/2\rangle$ energy level.
The presence of the magnetic field along the quantization axis is taken into account and ensures that the $|27G_{9/2}, m_J=9/2\rangle$ energy level is well separated from the neighboring levels as shown in Fig.~\ref{figStark}. The state $|27G_{9/2}, m_J=9/2\rangle$ corresponds to the state $|n=27, L=4, S=1/2, m_L=4, mS=1/2\rangle$ which is a non-degenerate eigenstate of the atom in the magnetic field.
Note that the fine structure splitting of the $27G$ multiplicity is neglected since a measurement of ours shows that it is smaller than 0.5~MHz, which is expected as explained in Ref.~\cite{lee2016microwave}.
\begin{figure}[t!]
\begin{center}
\includegraphics[clip,width=7.5cm]{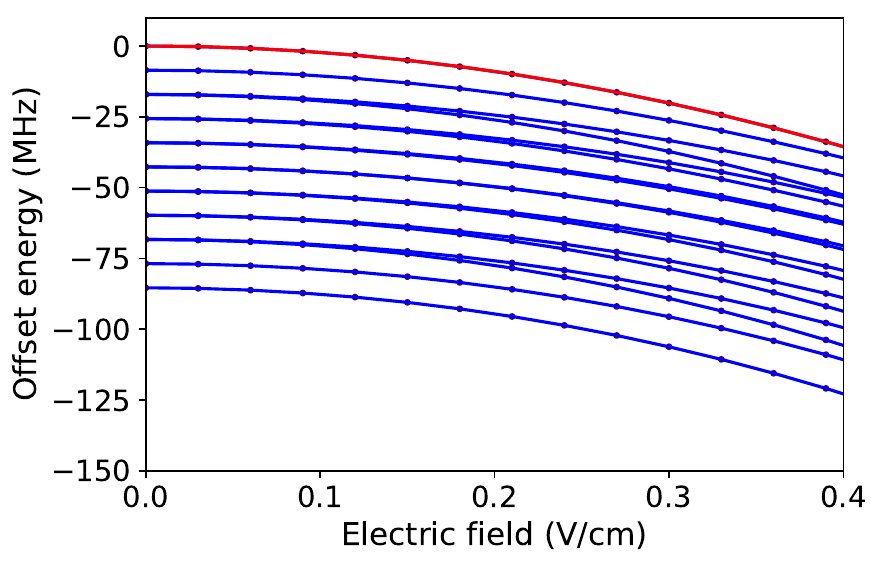}
\end{center}
\caption{\label{figStark}
Stark diagram of the $27G$ multiplicity in the configuration  of perpendicular electrostatic and magnetic field, with the magnetic field aligned along the quantization axis. The energy is offset from that of $|27G_{9/2}, m_J=9/2\rangle$. The level highlighted in red is $|27G_{9/2}, m_J=9/2\rangle$.
}
\end{figure}
\begin{figure}[t!]
\begin{center}
\includegraphics[clip,width=7.5cm]{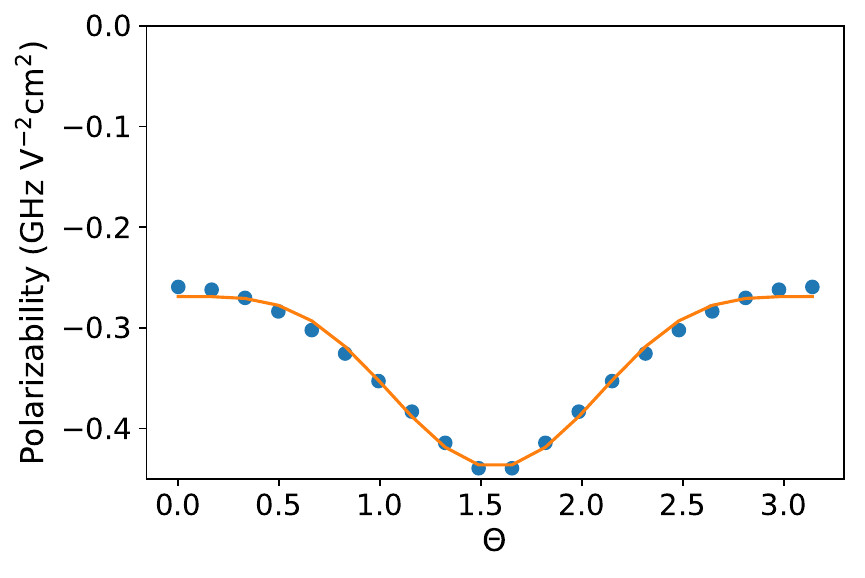}
\end{center}
\caption{\label{polarizability}
Polarizability of the $|27G_{9/2}, m_J=9/2\rangle$ energy level as a function of the polar angle $\theta$ between the quantization axis and the local axis of the electrostatic field. The red line is a fit using Eq. \eqref{polarEq} to the numerical data (blue circles).
}
\end{figure}
%
\subsection{Calculation of the probe light transmission}
The transmission of the probe light is given by
\begin{equation}
T(x,y)=\left|\frac{\Omega_P (x,y,+\infty)}{\Omega_P (x,y,-\infty)} \right|^2
\label{transmission}
\end{equation}
 for some transverse position $(x,y)$ and is calculated after integrating Eq. \eqref{maxwell} along $z$ using the Euler's integration method.
To calculate the actual transmission in some region of interest, $T(x,y)$ is computed on a square grid with spacing equal to the pixel size of the image in the experiment (2.54 $\mu$m).

\subsection{The transmission in the presence of expansion and Programming sequence}

For each ion cloud configuration, we calculate an initial random distribution of ions and its expansion as defined in subsection \ref{expansion}.
While the expansion is neglected during the ionization duration of $0.14\ \mu$s, it is taken into account during the $0.1\ \mu$s delay time between excitation and detection and during the imaging exposure time $\Delta t=1 \ \mu$s. The delay time has barely any effect on the overall expansion of the ion cloud for the ion densities that we consider in the experiment, whereas the expansion during the detecion stage is not negligible for ion numbers larger than about ten.

The imaging time $\Delta t$ is divided into several time steps $\delta t$ during which the expansion of the ion cloud is sufficiently small to assume that the transmission of the probe field is constant.
In practice, we calculate the expansion configuration of the ion cloud at time $t$, then calculate the transmission with formula \eqref{transmission} using the energy shifts determined by this configuration, before moving to the next time step $t+\delta t$.
The final averaged transmission is computed with the Trapezoidal method of integration.
The detection exposure time of $\Delta t$ is typically divided into $\sim 20$ steps. For each varied experimental parameter (for example $\Delta_P$), we then calculate the average transmission in some given region of interest over many ion cloud configurations ($\sim 300-800$).

\section{The scaling of the transmission with the number of ions}
A region of interest (ROI) is chosen as a part of the image of the ion cloud. In the absence of ions, an average of $N_{EIT}$ probe light photons are transmitted within the ROI during the imaging time $\Delta t$. With $P_{0}$ the incoming probe light power and $\overline{T}_0$ the average transmission within the ROI, $N_{EIT}$ is simply given by

\begin{equation}
N_{EIT}=P_{0} \ \overline{T}_0 \times \Delta t .
\end{equation}
The effect of one ion is to block the excitation to Rydberg state within a blockade distance $R_b$ from the ion, which scales as $\alpha^{1/4}$, where the polarizability $\alpha$ depends on the polar angle $\theta$. The angular dependence of $\alpha$ defines an oblate blockade surface, with the largest blockade distance occurring in the configuration of perpendicular electrostatic and magnetic fields. Note here that the direction of the electrostatic field is along the ion-atom interatomic axis. Within the volume enclosed by the blockade surface, the absorption of the probe light by the atoms is approximately that of two-level atoms. Here, the two-level absorption curve depends on the off-resonant D field.

Given a dilute cloud of ions, the blockade volumes surrounding the ions can be considered to be independent of each others if no pair of ions  $\left\{ \overrightarrow{r_i}, \overrightarrow{r_j} \right\}$ is such that $\sqrt{(x_j-x_i)^2+(y_j-y_i)^2} < R_{b,\perp}$, where $R_{b,\perp}=R_{b}(\theta=\pi/2)$. This approximation leads to the following expression for the number of transmitted photons in presence of ions
\begin{equation}
N_{ph}= N_{EIT}- \sum_j N_{sub,j},
\label{TheTotalPhotonNumber}
\end{equation}
where $N_{sub,j}$ is the photon number subtracted from $N_{EIT}$ due to the presence of ion $j$ which is located at a random position in the ion cloud.
Averaging over many realizations we obtain the average number of transmitted photons in presence of the ion cloud as
\begin{equation}
N_{ph,av}\approx N_{EIT}- N_i N_{sub},
\label{TheTotalPhotonNumber}
\end{equation}
where $N_i$ is the number of ions and $N_{sub}$ is the mean number of photons subtracted from $N_{EIT}$ when only one ion is present in the ion cloud. Note that the smaller the ROI, the smaller $N_{sub}$ is. Since $N_{ph,av}$ relates to the average transmission within the ROI in presence of ions, $\overline{T}_i$, as $N_{ph,av}=P_{0} \ \overline{T}_i \times \Delta t$,
we obtain finally
\begin{equation}
\mc{A}=1-\overline{T}_i/\overline{T}_0 \approx N_i N_{sub}/N_{EIT}.
\end{equation}
For large $N_i$, the blockade surfaces intersect and the assumption of a simple collection of independent scattering volumes fails. The relative absorption $\mc{A}$ is no longer linear with $N_i$ and rapidly saturates.\\

\end{document}